\documentclass[12pt]{article}
\usepackage{amsmath, amsthm,comment}
\usepackage{amsfonts}
\usepackage{amssymb,graphics,psfrag}
\usepackage{array,epsfig,stmaryrd,graphicx}

\newcommand{\beq}{\begin{equation}}
\newcommand{\eeq}{\end{equation}}
\newcommand{\beqa}{\begin{eqnarray}}
\newcommand{\eeqa}{\end{eqnarray}}
\newcommand{\beqar}{\begin{eqnarray*}}
\newcommand{\eeqar}{\end{eqnarray*}}

\newcommand{\reef}[1]{(\ref{#1})}
\begin{document}
\baselineskip 18pt%
\begin{titlepage}
\vspace*{1mm}%
\hfill
\vbox{

    \halign{#\hfil         \cr
   %       hep-th/yymmnnn\cr
         CERN-PH-TH/2013-xyz\cr
        % IPM/P-2010/003  \cr
         %  CPHT RR-xxx .yyzz \cr
           } % end of \halign
      }  % end of \vbox
\vspace*{10mm}
\vspace*{12mm}%

\center{ {\bf \Large  More On Critical Collapse of Axion-Dilaton \\ System  in Dimension Four
% for Elliptic Ansatz,Parabolic case and Hyperbolic case

}}\vspace*{3mm} \centerline{{\Large {\bf  }}}
\vspace*{5mm}
\begin{center}
{Luis \'Alvarez-Gaum\'e $^{a}$ and  Ehsan Hatefi  $^{b}$}
%$\footnote{On leave of Ferdowsi University of Mashhad }$

\vspace*{0.8cm}{ {
Theory Group, Physics Department, CERN, CH-1211, Geneva 23, Switzerland $^{a}$\\
and\\
International Centre for Theoretical Physics,
 Strada Costiera 11, Trieste, Italy $^{b}$}}$\footnote{E-mails:Luis.Alvarez-Gaume,ehsan.hatefi@cern.ch,ehatefi@ictp.it}$
\vspace*{1.3cm}
\end{center}
\begin{center}{\bf Abstract}\end{center}
\begin{quote}

We complete our previous study of critical gravitational collapse in the axion-dilaton system by analysing the
hyperbolic and parabolic ans\"atze.  As could be expected, the corresponding Choptuik exponents in four-dimensions
differ from the elliptic case.
\end{quote}
\end{titlepage}%

 %%%%%%%%%%%%%%%%%%%%%%%%%%%%%%%%%%%%%%%%%%%%%%%%%%%%%%%%%%%%%%%%%%%%%%%%%% %%%%%%%%%%%%%%%%%%%%%%%%%%%%%%%%%%%%%%%%%%%%%%%%%%%%%%%%%%%%%%%%%%%%%%%%%%
\section{Introduction}

Choptuik scaling (\cite{Chop}, for reference on the relevant literature, see \cite{Gundlach:2002sx})
is a remarkable property in many cases of gravitational collapse that was discovered twenty years
ago.  A system that experiences gravitational collapse is the axion-dilaton system motivated in
part by String Theory.  The study of Choptuik scaling in such a system was started in
\cite{HE,Hamade:1995jx,Eardley:1995ns}.  In this paper we complete the extension of these papers started
in \cite{AlvarezGaume:2011rk}, where we considered the elliptic case (see section 2) in four and
higher dimensions.  In that paper we also argued the existence of two more possibilities to satisfy the conditions
of continuous self-similarity, but we did not perform the relevant computations to determine
their corresponding critical exponents.  This is what we do in this paper in the special case
of four-dimensions, just to show that criticality also appears in those cases.
\vskip.1in
This paper is organized as follows. For convenience we briefly describe the axion/dilaton system
and the different continuous self-similar ansatz\"e in section two, where we also write down
the equations of motion in four-dimensions and the initial conditions in each case. In section
three we present the critical solutions in the hyperbolic and parabolic cases in some detail,
and finally in section four we carry out the necessary perturbations to obtain the Choptuik
exponent in each case.  A number of appendices are dedicated to some technical details needed
in the numerical analysis.  More details and references can be found in \cite{AlvarezGaume:2011rk}.

\section{The axion/dilaton system}

The axion $a$ and dilaton $\phi$ field can be combined into a single complex field
$\tau \equiv a + i e^{- \phi}$, its dynamics and coupling to gravity is described by the action:
\begin{equation}
S = \int d^4 x \sqrt{-g} \left( R - {1 \over 2} { \partial_a \tau
\partial^a \bar{\tau} \over (\mathop{\rm Im}\tau)^2} \right) \; .
\label{eaction}\end{equation}
where $R$ is the scalar curvature. The equations of motion are:
\begin{eqnarray}
\label{eoms}
R_{ab} - {1 \over 4 (\mathop{\rm Im}\tau)^2} ( \partial_a \tau \partial_b
\bar{\tau} + \partial_a \bar{\tau} \partial_b \tau) & = & 0 \end{eqnarray}
\begin{eqnarray}
\label{eoms14}
\nabla^a \nabla_a \tau + { i \nabla^a \tau \nabla_a \tau \over
\mathop{\rm Im}\tau} = 0 .
\end{eqnarray}
As in our previous work \cite{AlvarezGaume:2011rk} we look for critical solutions by assuming
spherical symmetry and continuous self-similarity (CSS).
Following  \cite{HE,Hamade:1995jx,Eardley:1995ns} the metric is taken to be:
\begin{equation}
	ds^2 = \left(1+u(t,r)\right)\left(- b(t,r)^2dt^2 + dr^2\right)
			+ r^2d\Omega^2 \;.
\label{metric1}
\end{equation}
We define a scale invariant variable as $z \equiv - r/t$ so CSS means that  the dimensionless
functions $u(t,r),b(t,r)$
in the metric are expressed only in terms of $z$:  $b(t,r) = b(z),\, u(t,r) = u(z)$.

The CSS condition for $\tau$  was considered in
detail in \cite{AlvarezGaume:2011rk}. The axion-dilaton Lagrangian
has a global $SL(2,R)$-symmetry (it is broken to a $SL(2,Z)$-subgroup by
non-perturbative phenomena, see for instance \cite{SenRev2})
hence we can compensate the action of the homothety vector field
 ($\xi = t \frac{\partial}{\partial t} + r \frac{\partial}{\partial r}$)
by means of an $SL(2,R)$-transformation.  There are in fact three possible
ansatz\"e, depending on whether the $SL(2,R)$-transformation chosen is
elliptic, hyperbolic or parabolic. The elliptic ansatz is given by:
\begin{equation}
 \tau(t,r)	=  i { 1 - (-t)^{i \omega} f(z) \over 1 + (-t)^{i
\omega} f(z)} ,
\label{tauansatz}
\end{equation}
where $\omega$ is a real constant determined by regularity conditions of the critical solution\footnote{Its value cannot
vanish $\omega=0$ otherwise we just get the trivial solution $f(z)=constant,b(z)=1$.} The
critical solutions and their exponents were explored in \cite{AlvarezGaume:2011rk} from four to ten dimensions.

The second ansatz is called hyperbolic and is given by:
\begin{equation}
\tau(t,r) = \;\frac{1- (-t)^{\omega} f(z)}{1+ (-t)^{\omega} f(z)},
\label{tau2ansatz}
\end{equation}
 finally the parabolic ansatz is simply:  $\tau(t,r) = f(z)+\omega \log(-t)$.

\subsection{Equations of motion and initial conditions in  four dimension }

In \cite{AlvarezGaume:2011rk} one can find the equations of motion in  dimensions four up to ten  for all three ans\"atze.
Here for convenience we reproduce them  for the hyperbolic and parabolic  cases in four dimension.
Having used spherical symmetry, there are no  gravitational degrees of freedom left, thus $u(z),b(z)$ should be re-expressed just in terms of $f(z)$. If we use the Einstein equations for the angular variables  we obtain:

\begin{equation}
u(z)\,=\,-{z\, b'(z)\over \,b(z)}.
\end{equation}

The other equations of motion include $b(z), f(z)$ but the equation for $b(z)$ can be used
to express this function in terms of $f(z),f'(z)$.The time coordinate  is chosen so that  the gravitational
collapse takes place at $t=0$.  Using the scaling properties of the metric   \reef{metric1} we can set
regular  under time scaling  so that by making use of  this invariance
$b(t,0)=1$ for $t<0$ .  Regularity at the origin also implies $u(t,0)=0$ .

\vskip.2in

The equations of motion for the four dimension  for hyperbolic case are:
\begin{eqnarray}
0 & = & b' - { z(b^2 - z^2) \over b (f - \bar{f})^2} f' \bar{f}' +{
 \omega (b^2 - z^2) \over b (f - \bar{f})^2} (f \bar{f}' + \bar{f} f')
+ {\omega^2 z |f|^2 \over b (f - \bar{f})^2} \; \label{a7}\\
0 & = & -f''
     - {z (b^2 + z^2) \over b^2 (f - \bar{f})^2} f'^2 \bar{f}'
     + {2 \over (f - \bar{f})} \left(\frac{1}{\bar{f}}
       + { \omega (b^2 + z^2) \over 2 b^2 (f - \bar{f})} \right) \bar{f} f'^2
     + { \omega (b^2 + 2 z^2) \over b^2 (f - \bar{f})^2} f f'
\bar{f}' \nonumber \\
 && + {2 \over z} \left(-1 + { \omega z^2 (f +\bar{f})\over (b^2 - z^2)
(f - \bar{f})} + {\omega^2 z^4 |f|^2 \over b^2 (b^2 - z^2)(f - \bar{f})^2 }\right) f'- {\omega^2 z \over b^2
(f - \bar{f})^2} f^2
\bar{f}' \nonumber \\
&& + { \omega \over (b^2 - z^2)} \left(-1 - { \omega (f + \bar{f})
\over (f - \bar{f})} - {\omega^2 z^2 |f|^2 \over b^2 (f - \bar{f})^2}
\right) f \; \label{a8}.
\end{eqnarray}

We find it convenient to represent split $f(z)$ into its real and imaginary parts:
\begin{eqnarray}
f(z)=u(z)+iv(z).
\nonumber\end{eqnarray}

These equations are invariant under a constant scaling $f\rightarrow \lambda f$, therefore we have freedom to choose the  real value of $f(z)$ ($u(z)$) or its imaginary part ($v(z)$) as we wish at a particular value of $z$.
We use this freedom to set $u(0)=1$.  Requiring regularity at the origin
we find that $f'(z=0)$ should vanish. Thus  the initial conditions for the hyperbolic case are:
\begin{eqnarray}
b(0)=u(0)=1, u'(0)=v'(0)=0
\nonumber\end{eqnarray}
\vskip.2in

 For the parabolic case the equations of motion in four dimensions are:
\begin{eqnarray}
0 & = & b' +{ z( z^2-b^2) \over b (f - \bar{f})^2} {f}'\bar{f}' -{
 \omega ( z^2-b^2 ) \over b (f - \bar{f})^2} (\bar{f}' + f')
+ {\omega^2 z \over b (f - \bar{f})^2} \; \label{a20}\\
0 & = & -f''
     - {z (b^2 + z^2) \over b^2 (f - \bar{f})^2} f'^2 \bar{f}'
     + {2 \over (f - \bar{f})} \left(1
       + { \omega (b^2 + z^2) \over 2 b^2 (f - \bar{f})} \right) f'^2
     + { \omega (b^2 + 2 z^2) \over b^2 (f - \bar{f})^2} f'
\bar{f}' \nonumber \\
 && + {2 \over z} \left(-1 + { 2\omega z^2 \over (b^2 - z^2)
(f - \bar{f})} + {\omega^2 z^4 \over b^2 (b^2 - z^2)(f - \bar{f})^2 }\right) f'- {\omega^2 z \over b^2
(f - \bar{f})^2}
\bar{f}' \nonumber \\
&& + { \omega \over (b^2 - z^2)} \left(-1 - { 2\omega
\over (f - \bar{f})} - {\omega^2 z^2  \over b^2 (f - \bar{f})^2}
\right)\; \label{a21}.
\end{eqnarray}

we write again $f(z)=u(z)+i v(z)$ these equations are independent of $u(z)$, and a similar analysis to
the the previous case leads to the initial conditions:
\begin{eqnarray}
b(0)=1, u'(0)=v'(0)=u(0)=0.
\nonumber\end{eqnarray}
In this case, the equations of motion are invariant under shifts of $f(z)$ by a real number.  These conditions
(and those of the hyperbolic case) can also be imposed in any number of dimensions to look for critical solutions.

In the elliptic case, \cite{AlvarezGaume:2011rk}
writing $f(z) = f_m(z) e^{if_a(z)}$, the regularity conditions imply:
\begin{eqnarray}
b(0)=1,f_{m}'(0)=f_{a}'(0)=f_{a}(0)=0\label{bca}
\end{eqnarray}

\section{Properties of the critical solutions in the hyperbolic and parabolic cases  in four dimension }

In this section  we discuss briefly the properties of the critical hyperbolic and parabolic solutions.
We follow \cite{Hamade:1995jx,Eardley:1995ns}.

  In \reef{a7},\reef{a8},\reef{a20},\reef{a21} we do have five singular points, $z=\pm 0$ corresponds to origin of polar
coordinates where we have studied the regularity conditions.
The point $z=\infty$ describes the surface $t=0$. The easiest way to study the neighbourhood of   $z=\infty$
is to use a change of variables and a redefinition of the fields $f(z),b(z)$ \cite{Eardley:1995ns}.
This is discussed in detail in the Appendices.
\vskip.1in
The singularities $b(z_{\pm})=\pm z_{\pm}$ correspond to the surfaces where the
homothetic Killing vector becomes  null.  They are related to the backward
(forward) light cones of the space-time origin. For $b(z_+)=z_+$ the solution should be
smooth across this surface. However the forward cone $b(z_-)=-z_-$ represents the Cauchy horizon
of the space-time and we should not require more than continuity of $f,b$ in this region.
We need to impose smoothness of the space-time just a bit below the forward
cone and then extend the solution by continuity.

\vskip.1in
Using regularity at the origin and at $z_+$ and using the initial conditions described,
the critical solution is determined by the following four parameters (in both hyperbolic and
parabolic cases):
\begin{eqnarray}
 |v(0)|,\omega,z_+, |v(z_+)|
 \nonumber
 \end{eqnarray}
\vskip.1in
We explain our procedure to determine these parameters in the hyperbolic case in detail,
and just give the results in the parabolic case.

Taylor expanding at the origin, and using the regularity conditions we obtain:
\begin{eqnarray}
 b(z)& = &1+ \frac{\omega^2(1+v(0)^2)}{12(v(0)^2)}z^2+ O(z^{4})
 \nonumber
 \end{eqnarray}
\begin{eqnarray}
v(z)& = &v(0)+\frac{\omega(\omega-v(0)^2) }{3v(0) }z^2+ O(z^{4})
\nonumber
\end{eqnarray}
\begin{eqnarray}
u(z)& = &1+\frac{-\omega(1+\omega)}{3} z^2+ O(z^{4})
\end{eqnarray}

To obtain the remaining values of $\omega, v(0)$, we integrate out from the origin to positive values
of $z$ and we also integrate in from $z_+$ towards the origin.  We also Taylor expand at $z_+$ and
impose regularity.  Matching the two solutions at an intermediate point and requesting continuity
in the functions and their first derivatives, completely determines all parameters in the critical
solution.  The solution for hyperbolic is  given by the parameters:
\begin{eqnarray}
\omega & = & 1.127 ,\nonumber\\
z_{ +} & = & 1.561 ,\nonumber\\
|v(0_{ +})| & = & 0.397,\nonumber\\
|v(z_{ +})| & = & 1.016.
\label{last}
\end{eqnarray}
In the parabolic case we obtain:
\begin{eqnarray}
\omega & = & 1.200 ,\nonumber\\
z_{ +} & = & 1.601 ,\nonumber\\
|v(0_{ +})| & = & 0.321,\nonumber\\
|v(z_{ +})| & = & 1.071
\label{last}\end{eqnarray}
We used rather moderate precision to determine the parameters in the solution.
If one wants to determine $z_-,f(z_-)$ as well, the precision has to be increased substantially
as a consequence of the fact that the solution is relatively flat in the forward light-cone.
Some details are given in Appendix A in the elliptic case.
\vskip.1in

\section{Choptuik exponent in the hyperbolic case}

We follow the standard methods to compute the critical exponent
\cite{Gundlach:2002sx, Hamade:1995jx,AlvarezGaume:2011rk}. We perturb the critical solution:
\begin{equation}
f(z,t) = f_{\rm ss}(z) + \epsilon |t|^{- \kappa} f_{\rm pert}(z),
  \nonumber\end{equation}
to obtain a set of linear equations  for the perturbations with an
eigenvalue equation for  $\kappa$.
By taking  the biggest value for $Re(\kappa)$,
the critical exponent can be found as $\gamma = \frac{1}{Re(\kappa)}$.  The linear equations
have singular points in the same places as the critical solution.  Once regularity is imposed
we can determine completely the most relevant mode and the value of $\kappa$.
\vskip.1in

We give a few details in what follow on the method we use to derive the perturbation
equations.  First, we keep the general form of the metric in terms of the functions
$b(t,r)$ and $u(t,r)$ in the equations of motion.
In addition, we find it convenient to write for the axion-dilaton field:
\begin{equation}
\tau(t,r)= \frac{1-f(t,r)}{1+f(t,r)}
\nonumber\end{equation}
Then we introduce the variations of the fields $u(t,r),b(t,r)$ and $\tau(t,r)$, as follows:
\begin{equation}
\delta u(t,r)=(-t)^{-\kappa}\,u_{1}(t,r),\quad\quad\quad
\delta b(t,r)=(-t)^{-\kappa}\,b_{1}(t,r)
  \nonumber\end{equation}
\begin{equation}
\delta \tau(t,r)=\frac{-2\delta f(t,r)}{(1+f(t,r))^2},\quad\quad\quad \delta f(t,r)=(-t)^{\omega-\kappa}f_{1}(t,r),
  \nonumber\end{equation}
where $u_{1}(t,r),b_{1}(t,r)$ and  $f_{1}(t,r)$ are the perturbations.  The CSS ansatz for the critical solution
and the perturbations are:
\begin{equation}
b(t, r) = b(-r/t),\quad\quad u(t, r) = u(-r/t)\nonumber\end{equation}
\begin{equation}
\delta u(t, r)= (-t)^{-\kappa} u_{1}(-r/t),\quad\quad \delta b(t, r)= (-t)^{-\kappa} b_{1}(-r/t)\nonumber\end{equation}
\begin{equation}
 \tau(t,r)	=   { 1 - (-t)^{ \omega} f(-r/t) \over 1 + (-t)^{\omega} f(-r/t)} ,\quad\quad \delta\tau(t,r)=-2 {  (-t)^{ \omega-\kappa} f_{1}(-r/t) \over (1 + (-t)^{\omega} f(-r/t))^2}.\nonumber
\end{equation}
The perturbed equations are rather cumbersome, and there is not much point in spelling them out explicitly here.  Some
details appear in Appendix D.  As with the critical solution, some algebraic manipulations allow us to determine
$u_{1},u'_{1},u,u',$  in terms of $b_1$ and $f_1$.
\vskip.2in
As in the critical case, $b_1(z)$ is determined in terms of $f_1(z)$ and the critical solution.  The linear equation
for $f_1(z)$ has the same singularities as the the unperturbed equations at $z=0,z=\infty$ and at $z_{\pm}$.

We do require that the perturbations be smooth at the singularities of
the original equations and this provides  all the perturbations  up to some re scalings.

Thus we found a set of linear equations for
the perturbations which have solutions for some values of $\kappa$. The solution with the biggest value of
$Re (\kappa)$ corresponds to Choptuik exponent.  As usual:
\begin{eqnarray}
\nonumber\gamma = \frac{1}{Re(k)}.
\end{eqnarray}
\vskip.1in
Therefore by analyzing the perturbations appeared in Appendix D, we obtain in the hyperbolic case
\begin{eqnarray}
\gamma=\,0.363
\end{eqnarray}

\section{Choptuik exponent in the parabolic case}

In this section we simply mention the differences with the previous case.
The ansatz for the critical solution and the perturbation are in this case:
\begin{equation}
b(t, r) = b(-r/t),\quad\quad u(t, r) = u(-r/t)\nonumber\end{equation}
\begin{equation}
\delta u(t, r)= (-t)^{-\kappa}\,u_{1}(-r/t),\quad\quad \delta b(t, r)= (-t)^{-\kappa}\, b_{1}(-r/t)\nonumber\end{equation}
\begin{equation}
 \tau(t,r)	=  f(-r/t) +  \omega  \log(-t)  ,\quad\quad \delta\tau(t,r)=(-t)^{-\kappa}\, f_{1}(-r/t).\nonumber
 \end{equation}
Following the same steps as in the hyperbolic case, we find the following value for the critical exponent:
\begin{eqnarray}
\gamma=0.324
\end{eqnarray}
Just for comparison, we recall that the value of $\gamma$ for the elliptic case is [4]
$\gamma=0.264$.

This shows clearly that the exponent in the three cases are different.
It is important to point out that it would be useful to use more powerful numerical methods
in order to get more accurate expressions for the exponents.  Our interests was to use enough
precision so as to distinguish the exponents in the parabolic, hyperbolic and  elliptic cases
\cite{AlvarezGaume:2011rk}.  As remarked in that paper, if instead of assuming CSS one
only requires discrete self-similarity, the discrete scale transformation can then be compensated
by an element of $SL(2,Z)\,\in\,SL(2,R)$.  It would be interesting to know how the critical
exponents depend on the modular transformations.

\section*{Acknowledgment}
E.H acknowledges the Theory Group at CERN for its hospitality while this work was being completed.
He also thanks E.Hirschmann for several valuable  communications.
%%%%%%%%%%%%%%%%%%%%%%%%%%%%%%%%%%%%%%%%%%%%%%%%%%%%%%%%%%%%%%%%%%%%%%%%%%%%%%%%%%%%%

%\section{Appendices}

 \section*{Appendix A : The change of variables at infinity and the solutions for the elliptic case}

Here we show how using field redefinitions we can study the equations at $z=\infty$.
Using the representation  $f(z)=f_{m}(z)e^{if_{a}(z)}$
we can divide e.o.m's to the modulus and phase parts, and find their leading behaviour
at infinity. By replacing  $f_{m}(z)=u z^{k}$ and $f_{a}(z)=v z^{s }$ in the equations of motion we
obtain near infinity: $f(z)=z^{-iw}e^{if_{0}}$. Thus the equations
of motion  for the elliptic case become regular there
by the following change of variables:
\begin{eqnarray}
dw  & = & b(z){dz\over z^2}, \quad w=0{\rm{\quad at}\quad}z=\infty \; ,\nonumber\\
   F(w)	& = & z^{-i \omega} f(z) \; ,\nonumber\\
   v(w)	& = & {b(z)\over z} \; ,			\nonumber\\
   u(w) & = & u(z) \; .
\label{k211}\end{eqnarray}

In four dimensions it is shown in \cite{Eardley:1995ns} how to analyze the equations of motion
in terms of new independent variable $w$.  It is straightforward to extend
their arguments to five dimensions, and we find:
\begin{eqnarray}
0 & = & v' + {2(v^2 - 1) \over 3(1 - |F|^2)^2} F' \bar{F}' + 1 - {2\omega^2
|F|^2 \over 3(1 - |F|^2)^2} \; , \label{10}\\
0 & = & F'' -  {2(2 v F'+ i \omega F) \over 3(1 - |F|^2)^2} F' \bar{F}' +
  {2 \bar{F} F'^2 \over 1 - |F|^2}
  + { v \over (v^2 - 1) } \bigg(1 +\frac{2 i \omega (1 + |F|^2)}{(1 - |F|^2)} +
  {4 \omega^2 |F|^2 \over 3(1 - |F|^2)^2} \bigg) F'
\nonumber \\
&&
- {i \omega \over v^2 -1 } \left( -2 - {i \omega (1 + |F|^2)
  \over 1 - |F|^2 } - {2\omega^2 |F|^2 \over 3(1 - |F|^2)^2}
  \right) F \; .
  \label{11}
  \end{eqnarray}

Near $ z_{-}$ the space-time becomes nearly flat
 so one has to increase the accuracy of the parameters.
To be able to distinguish $z_{-}$ in all  dimensions of space-time we have
to be more precise as we will see in the table 1. Therefore near $z_{-}$ we
increase the precision of the numerical computation.
Then we can use of the new coordinate $w,v,F$  (\ref{k211}) to integrate from
$w=w_{ +}$ until the next singularity at
$w=w_{ -}$. It is important to clarify that $F''$ and all higher derivatives of $F$ are discontinuous at
$w_{ -}$. This in principle physically acceptable.

\vskip.1in

To find the second singularity we need to take into account the region $0 > z > z_{ -}$.
Since $f$ and $f'$ are continuous at $z=z_{ -}$,
we start integration at  $z= - \epsilon_0$ and integrate out to $z_{ -}$. Time rescaling allows us to
impose once more $b(0_{-}) = 1$ and
requiring regularity  at $z = 0_{-}$ , we find the solutions in all ten
dimensions of space time for elliptic case   in table 1.

\begin{table}[h]
\centering
\begin{tabular}{ccccc}
\hline
$dimension$\quad\quad\quad $\omega$\quad\quad\quad\quad\quad\quad $z_{-}$\quad\quad\quad\quad\quad $|f(0_{-})|$\quad\quad\quad\quad $|f(z_{-})|$\quad \\
\hline
\quad\quad\quad4\quad\quad\quad\quad1.1769527\quad\quad$-1.0000372$\quad\quad.0117429\quad\quad$.0076030$\quad\quad\\
\\
\quad\quad\quad5\quad\quad\quad\quad 1.2976152\quad\quad$-1.0000302$\quad\quad.0100758\quad\quad$.0071049$\quad\quad\\
\\
\quad\quad\quad6\quad\quad\quad\quad1.4695887\quad\quad$-1.0000403$\quad\quad.0111984\quad\quad$.0081036$\quad\quad\\
\\
\quad\quad\quad7\quad\quad\quad\quad1.6107643\quad\quad$-1.0000450$\quad\quad.0116716\quad\quad$.0086207$\quad\quad\\
\\
\quad\quad\quad8\quad\quad\quad\quad1.7218682\quad\quad$-1.0000534$\quad\quad.0127263\quad\quad$.0095849$\quad\quad\\
\\
\quad\quad\quad9\quad\quad\quad\quad1.7910669\quad\quad$-1.0000563$\quad\quad.0132912\quad\quad$.0102488$\quad\quad\\
\\
\quad\quad\quad10\quad\quad\quad\quad1.8524703\quad\quad$-1.0000747$\quad\quad.0155778\quad\quad$.0122424$\quad\quad\\
\\
\hline
\end{tabular}
\caption{Finding solutions for the elliptic case at $z_{-}$ in dimensions $4,5,6,7,8,9,10 $, respectively.}
\label{qmax2}
\end{table}

\section*{Appendix B : The change of variable at infinity  for the hyperbolic case}

The analysis of $z=\infty$ can be extended for the hyperbolic and parabolic cases.
In the hyperbolic case, the equations of motion become  regular at $z=\infty$
through the change of variables :
\begin{eqnarray}
dw  & = & b(z){dz\over z^2}, \quad w=0{\rm{\quad at}\quad}z=\infty \; ,\nonumber\\
   F(w)	& = & z^{-\omega} f(z) \; ,\nonumber\\
   v(w)	& = & {b(z)\over z} \; ,			\nonumber\\
   u(w) & = & u(z) \; .
\label{kk}\end{eqnarray}
leading to the equations:
\begin{eqnarray}
0 & = & v' - {v^2 - 1 \over (F - \bar F)^2} F' \bar{F}' + 1 + {\omega^2
|F|^2 \over (F - \bar F)^2} \label{pp},\\
0 & = & -F'' -  {2 v F'+  \omega F \over (F - \bar F)^2} F' \bar{F}' +
  {2 F'^2 \over (F - \bar F)}
  + {2 \omega v \over (v^2 - 1)  (F - \bar F)} \left(  (F + \bar F) +
  { \omega |F|^2 \over  (F - \bar F)} \right) F'
\nonumber \\
&&
+ { \omega \over v^2 -1 } \left( -1 + { \omega  (F + \bar F)
  \over  (F - \bar F) } + {\omega^2 |F|^2 \over  (F - \bar F)^2}
  \right) F \; \label{tt}.
\end{eqnarray}

\section*{Appendix C : The change of variable at infinity  for the parabolic case}

In this case the change of variables near $z=\infty$ is:
\begin{eqnarray}
dw  & = & b(z){dz\over z^2}, \quad w=0{\rm{\quad at}\quad}z=\infty \; ,\nonumber\\
   F(w)	& = & f(z) \; ,\nonumber\\
   v(w)	& = & {b(z)\over z} \; ,		\nonumber\\
   u(w) & = & u(z) \; .
\end{eqnarray}
leading to following equations of motion in $d=4$:
\begin{eqnarray}
0 & = & v' +{v^2 - 1 \over (F - \bar F)^2} \left(-F' \bar{F}'+{\omega(F'+\bar{F}')\over v}\right) + 1 + {\omega^2
 \over v^2(F - \bar F)^2} \label{pp9},\\
0 & = & -F'' -  {2 v F' \over (F - \bar F)^2} F' \bar{F}' +
  {2 F' \over (F - \bar F)}\left(F'+{\omega (F'+\bar{F}')\over (F - \bar F)}\right)
  + {2 \omega \over (v^2 - 1)  (F - \bar F)} \bigg({2\over v}\nonumber\\&&+
  { \omega \over  v(F - \bar F)}\bigg)F'
+ { \omega \over v^2 -1 } \left( -{1\over v^2 }- { 2\omega
  \over v^2 (F - \bar F) } - {\omega^2 \over  v^4(F - \bar F)^2}
  \right) \nonumber\\&&-{\omega \over v^2(F - \bar F)^2} \left({\omega (F'+\bar{F}')\over v}-F' \bar{F}'\right) \; \label{tt86}.
\end{eqnarray}

\section*{Appendix D: Perturbed equations for hyperbolic and parabolic case in four dimension}

The perturbation equations used in the hyperbolic case in four-dimensions to get the Choptuik exponent are:
\begin{eqnarray}
b'_{1}(r)& = &L_{1}\bigg(b_{1}(r) (f(r)- \bar f(r))^3 b'(r)
(1+k+b'(r))-2 b(r)^2 b_{1}(r) (f(r)-\bar f(r)) (1+b'(r))\nonumber \\&&
 \times f'(r)\bar f'(r)
+wb(r) (1+b'(r)) \bigg[\bar f(r)^2 (( k-w) f_{1}(r)-2  b_{1}(r) f'(r))+f(r) \bar f(r)\nonumber \\&&
\times (-( k+w) (f_{1}(r)-\bar f_{1}(r))+2 b_{1}(r )( f'(r)-\bar f'(r)))
+f(r)^2[(- k+w) \bar f_{1}(r)+\nonumber \\&&+2  b_{1}(r)
\bar f'(r)]\bigg]\bigg)
\nonumber\end{eqnarray}

\begin{eqnarray}
f_{1}'(r)&=&L_{2}\bigg(b(r) ((k-w) f_{1}(r) \bar f(r)(1+k-w-b'(r))-w f(r)^2 b'_{1}(r)
+2b(r)(f_{1}(r)\nonumber \\&&\times(2 k- w+b'(r))-\bar f_{1}(r)(w+b'(r))-\bar f(r)b'_{1}(r))f'(r)+f(r)(-(k+k^2 \nonumber \\&&+2kw-2w  (1+w)) f_{1}(r)+(k-2w) f_{1}(r)b'(r)+w\bar f_{1}(r)(-1+w+b'(r)) \nonumber \\&&+b'_{1}(r)(w\bar f(r)+2b(r)f'(r))))
+b_{1}(r)(w(1-k+w)f(r)^2+b(r)(-4b(r)f'(r)^2\nonumber \\&&+\bar f(r)(-(4+k+2w+2 b'(r)) f'(r)-2b(r)f''(r)))+f(r)\nonumber \\&&\times
(w(-1+k+w)\bar f(r)+b(r)((4+k-2w+2b'(r)) f'(r)+2b(r)f''(r)))).\bigg)\nonumber
\end{eqnarray}
Similarly, in the parabolic case:
\begin{eqnarray}
b'_{1}(r)& = &L_{3}\bigg(b_{1}(r) (f(r)- \bar f(r))^3 b'(r)
(1+k+b'(r))-2 b(r)^2 b_{1}(r) (f(r)-\bar f(r)) (1+b'(r))\nonumber \\&&
 \times f'(r)\bar f'(r)
-wb(r) (1+b'(r)) \bigg[\bar f_{1}(r)(-2w+kf(r)-k\bar f(r))+f_{1}(r)(2w+kf(r)\nonumber \\&&-k\bar f(r))-2b_{1}(r) (f(r)- \bar f(r))(f'(r)+ \bar f'(r))\bigg]\bigg)
\nonumber
\end{eqnarray}

\begin{eqnarray}
f_{1}'(r)&=&L_{4}\bigg(k(1+k)b(r)f_{1}(r)-krb'(r)f_{1}(r)+b(r)b'(r)b_{1}(r)f'(r)-(b(r))^2b'_{1}(r)f'(r)\nonumber \\&&
+rb'_{1}(r)(w-rf'(r))+\frac{2wb(r)(f_{1}(r)-\bar f_{1}(r))(w-2b(r)f'(r))}{(f(r)-\bar f(r))^2}\nonumber \\&&+\frac{4kb(r)f_{1}(r)(w-b(r)f'(r))}{(f(r)-\bar f(r))}+\frac{4b_{1}(r)(w-b(r)f'(r))^2}{(f(r)-\bar f(r))}+b_{1}(r)(-w+kw\nonumber \\&&+(2-k) rf'(r)+r^2f''(r))
-3b_{1}(r)(-w+wb'(r)-b(r)(-2+b'(r))f'(r)\nonumber \\&&+(b(r))^2f''(r))\bigg)\nonumber\end{eqnarray}

where $L_{1},L_2,L_{3},L_4$ are defined as:
\begin{equation}
L_{1}=\frac{1}{b(r)(f(r) -\bar f(r))^3(-1+k-b'(r))},\nonumber\end{equation}
\begin{equation}
L_{2}=\frac{1}{2 b(r)^2 \bigg(f(r) (k+ w-b'(r))+
\bar f(r) (-k+ w+b'(r))\bigg)}\nonumber\end{equation}

\beqa
L_{3}=\frac{1}{b(r)(f(r) -\bar f(r))^3(-1+k-b'(r))}\nonumber\eeqa
\beqa
L_{4}=\frac{1}{\bigg(-2(1+k)rb(r) +2\frac{b(r)^3}{r}+b(r)^2(\frac{4w}{(-f(r)+\bar f(r))}+b'(r))+r^2b'(r)\bigg)}\nonumber
\eeqa
 \vskip.1in
where  $r$ in the definitions of $L_i, i=1,\ldots$ is  $z_+$.


\begin{thebibliography}{2007}
\bibitem{Chop} M.W.~Choptuik,
{ Universality and Scaling in Gravitational Collapse of a Massless Scalar Field,}
{\it Phys.\ Rev. Lett.}\ {\bf70}, 9 (1993).
 \bibitem{Gundlach:2002sx}
  C.~Gundlach,
  % ``Critical phenomena in gravitational collapse,''
  Phys.\ Rept.\  {\bf 376}, 339 (2003)
  [gr-qc/0210101].
\bibitem{HE} E.W.~Hirschmann and D.M.~Eardley,
{ Universal Scaling and Echoing in Gravitational Collapse of a Complex Scalar Field,}
{\it Phys.\ Rev.}\ {\bf D51}, 4198 (1995), gr-qc/9412066.
%\cite{Hamade:1995jx}
\bibitem{Hamade:1995jx}
  R.~S.~Hamade, J.~H.~Horne and J.~M.~Stewart,
  ``Continuous Self-Similarity and $S$-Duality,''
  Class.\ Quant.\ Grav.\  {\bf 13} (1996) 2241
  [arXiv:gr-qc/9511024].
  %%CITATION = CQGRD,13,2241;%%
\bibitem{Eardley:1995ns}
%\cite{Eardley:1995ns}
  D.~M.~Eardley, E.~W.~Hirschmann and J.~H.~Horne,
``S duality at the black hole threshold in gravitational collapse,''
  Phys.\ Rev.\  D {\bf 52} (1995) 5397
  [arXiv:gr-qc/9505041].
  %%CITATION = PHRVA,D52,5397;%%
\bibitem{AlvarezGaume:2011rk}
  L.~Alvarez-Gaume and E.~Hatefi,
   ``Critical Collapse in the Axion-Dilaton System in Diverse Dimensions,''
  Class.\ Quant.\ Grav.\  {\bf 29}, 025006 (2012)
  [arXiv:1108.0078 [gr-qc]].
  %%CITATION = ARXIV:1108.0078;%%
%\cite{SenRev2}
\bibitem{SenRev2} Useful reviews are
  A.~Sen,``Strong - weak coupling duality in four-dimensional string theory,''
  Int.\ J.\ Mod.\ Phys.\ A {\bf 9}, 3707 (1994)
  [hep-th/9402002];
J.H.~Schwarz,
``Evidence for nonperturbative string symmetries,''
Lett.\ Math.\ Phys.\  {\bf 34}, 309 (1995)
  [hep-th/9411178].
\end{thebibliography}
\end{document}